\newcommand{\R}{R}
\newcommand{\gd}{\dot{\gamma}} 

\documentstyle[multicol,aps,psfig]{revtex}

\begin{document}

\draft

\title{Rheological Chaos in a Scalar Shear-Thickening Model}

\author{M. E. Cates,$^{1}$ D. A. Head,$^{1,2}$ and
A. Ajdari$^{3}$}

\address{$^{1}$Department of Physics and Astronomy, JCMB King's Buildings,
University of Edinburgh, Edinburgh EH9 3JZ, Scotland\\
$^{2}$Division of Physics and Astronomy,
Vrije Universiteit, De Boelelaan 1081, NL-1081 HV
Amsterdam, The Netherlands\\
$^{3}$Laboratoire de Physico--Chimie Th\'eorique, UMR CNRS 7083,
ESPCI, 10 rue Vauquelin, F-75231 Paris Cedex 05, France}

\date{\today}

\maketitle

\begin{abstract}
We study a simple scalar constitutive equation for a
shear-thickening material at zero Reynolds number,
in which the shear stress $\sigma$ is driven at a
constant shear rate $\dot\gamma$ and relaxes by
two parallel decay processes: a nonlinear decay at
a nonmonotonic rate $R(\sigma_1)$ and a linear
decay at rate $\lambda\sigma_2$.
Here $\sigma_{1,2}(t) = \tau_{1,2}^{-1}\int_0^t\sigma(t')\exp[-(t-t')/\tau_{1,2}]\,{\rm d}t'$ are two retarded stresses. For suitable parameters, the steady state flow curve is monotonic but unstable; this arises when $\tau_2>\tau_1$ and $0>R'(\sigma)>-\lambda$ so that monotonicity is restored only through the strongly retarded term (which might model a slow evolution of material structure under stress). Within the unstable region we find a period-doubling sequence leading to chaos. Instability, but not chaos, persists even for the case $\tau_1\to 0$. A similar generic mechanism might also arise in shear thinning systems and in some banded flows.
\end{abstract}



\maketitle

\begin{multicols}{2}

\narrowtext

Rheochaos can be defined as the occurence of macroscopic chaos \cite{Chaosfoot} in a viscoelastic material at negligible Reynolds number. With the neglect of inertia that this implies, the nonlinearity must come not from the advection of momentum (as in Navier Stokes turbulence) but from the constitutive behaviour of the material, which may include strong memory effects. Likewise for the chaos to be macroscopically observable (for example in time series data on the stress measured at fixed strain rate, or vice versa, in a bulk sample) a mechanism must be present that goes beyond the microscale chaos known to be present in, e.g., colloidal Stokes flow \cite{Chaikin}.

Strong candidates for rheochaos include micellar materials \cite{Sood1}, dense lamellar phases \cite{Jacques}, and also dense suspensions where erratic stress response at fixed strain rate (or vice versa) is widespread but poorly documented (see e.g. \cite{Laun}). It is not yet clear whether spatial as well as temporal inhomogeneity is present for all instances of rheochaos, and if so to what extent. This could range from a shear-banded flow in which the interface between bands of fast and slow flowing material is unsteady in time (as suspected in micelles \cite{Sood1,Callaghan}) through to fully developed `elastic turbulence' as recently reported in polymer solutions near the overlap threshold \cite{Groisman}.
Spatial inhomogeneities are also known to occur in
shear--thickening colloid solutions \cite{Laun,Newstein}.
However, the closely related phenomenon of director chaos in sheared nematics has been studied theoretically and does not seem to require spatial inhomogeneity\cite{Keunings}. In the present state of understanding, a theoretical search for temporal rheochaos in spatially homogenous models remains justified.

Recent work by the authors has studied the onset of temporal instability in spatially homogeneous mesoscopic models of shear-thickening type \cite{Head1}. One interesting prediction was that such instability could arise in a system where the steady state flow curve, $\sigma(\dot\gamma)$ is monotonic \cite{Head1}. This contrasts with the conventional instability to spatial inhomogeneity in the form of shear bands: this is always associated with regions of negative slope on the flow curve \cite{Ball,Spenley,Olmsted}. The mesoscopic models of \cite{Head1} are not fully tensorial but work with a single (spatially uniform) component of each of the stress and strain rate tensors ($\sigma$ and $\dot\gamma$); nonetheless they contain an infinite number of degrees of freedom, corresponding to the distribution of local strain variables for different mesoscopic elements. This makes them complex to analyse.

In this paper we propose closely related but much simpler models in which there is only one degree of freedom (the shear stress $\sigma$) whose time evolution at constant strain rate $\dot\gamma$ is governed by a simple constitutive equation with retarded and nonlinear features. The simplest such model combines a nonlinear instantaneous relaxation rate for stress (chosen nonmonotonic) with a linear but retarded relaxation. For a single exponential retardation kernel, its dynamics can be completely understood: it shows spontaneous oscillation in a region of the flow curve with positive slope, but no chaos. This is qualitatively very like the mesoscopic model of Ref.\cite{Head1}
(although that model exhibits oscillations at a constant imposed stress rather than strain rate). In particular, the instability is associated with a negative slope on the `bare' flow curve (before the retarded term is added). A second, similar model, in which the nonlinear relaxation is itself delayed, shows chaos.

We first examine the simplest model alluded to above. This is defined by the equation

\begin{equation}
\dot\sigma(t) = \dot\gamma - R(\sigma) -\lambda\sigma_2
\label{one}
\end{equation}

\noindent{}where
$\sigma_2(t) = \int_{-\infty}^t M_2(t-t')\sigma(t')\,dt'$ is a retarded stress and $M_2(t)$ is a memory kernel whose integral is unity. The first term on the right side of this equation means that, in the absence of relaxation, stress increases linearly with straining (the elastic constant is set to unity) -- a Hookean solid. The second term describes instantaneous decay of stress at rate $R(\sigma)$, for example through `hops' or plastic rearrangement of mesoscopic elements (returning these to an unstrained state) with jump rate $R/\sigma$. Unlike in the mesoscopic models of \cite{Head1}, no attempt is made to track the dynamics of individual elements. The third term is also a decay term, but describes retarded relaxation. This could represent `delayed jumps' which, perhaps because they involve a cooperative motion of many elements, take a distribution of finite times to accomplish (governed by the kernel $M_2$). More generally, a retarded term could represent some other slow structural reorganization of the material in response to stress.

For example, one could have a model of instantaneous jumps but with a `fluidity' or jump rate that itself adapts slowly to stress\cite{Derec}. In this context it might be more natural to have a nonlinear retarded term such as 

\begin{equation}
\dot\sigma = \gd - R(\sigma) -\lambda\sigma_2\sigma
\end{equation}

\noindent{}However, this gives qualitatively the same instability as described below for Eq.(\ref{one})~\cite{footnote}; we retain the linear version, for simplicity, below.

Solving Eq.(\ref{one}) in steady state gives immediately the flow curve, or rather its inverse:

\begin{equation}
\gd = \R(\sigma) + \lambda\sigma
\label{two}
\end{equation} 

\noindent{}The interesting case is
when $R(\sigma)$ is nonmonotonic but $R(\sigma)+\lambda\sigma$ is monotonic. Then the flow curve is monotonic, but only because of the retarded contribution to the jump rate. One might suspect that a sufficiently sluggish retarded contribution might fail to correct the underlying instability in the region where $R'(\sigma)$ is negative: over short timescales the system appears to unstable with respect to shear banding but at long timescales it is not. 
Here, the timescales are measured relative to the strain rate
at which $R(\sigma)$ in (\ref{two}) first becomes non--monotonic;
we choose units so that this is $O(1)$.

We analyse the case of a single exponential kernel,
$M_2 = \tau_{2}^{-1}\exp[-(t-t')/\tau_2]$. As is easily checked, for this kernel Eq.(\ref{one}) can be replaced by a differential equation of second order. Differentiating Eq.(\ref{one}) with respect to $t$, and noting that
$\dot\sigma_2 = (\sigma - \sigma_2)/\tau_2$,
we obtain immediately

\begin{equation}
\ddot\sigma = - \frac{\partial V}{\partial \sigma} -\xi(\sigma)\dot\sigma
\label{three}\end{equation}

\noindent{}which effectively describes a particle
of unit mass in a 1-D potential $V$ with
damping constant $\xi$. Here

\begin{eqnarray}
\tau_{2}\,V(\sigma) & = & \int_{0}^{\sigma} R(\sigma\prime)\, d\sigma\prime + \lambda\sigma^2/2 - \dot\gamma\sigma
\label{four}\\
\xi(\sigma) & = & R'(\sigma)+1/\tau_2 \label{five}
\end{eqnarray}

\noindent{}As $\dot\gamma$ is varied, the steady state flow curve $\sigma(\dot\gamma)$, as given by Eq.(\ref{two}), is recovered as the solution of $V'(\sigma) = 0$. Stability of the steady state solution requires that two further conditions are satisfied. The first is $V''(\sigma) > 0$ (so that the effective potential has a minimum not a maximum). This is equivalent to $d\sigma/d\dot\gamma > 0$ which is the usual criterion to avoid shear banding. However, stability also requires  that $\xi(\sigma)$ is positive at the minimum of $V$. When $R'(\sigma)$ in Eq.(\ref{five}) is negative, this is only satisfied if the retardation time $\tau_2$ is sufficiently short. When not satisfied, one has antidamping at the minimum of $V$ so that small velocity fluctuations are amplified; this is reminiscent of a van der Pol oscillator \cite{Glendinning}. Velocity fluctuations will grow until a limit cycle is reached in which the postive damping at large amplitudes balances the antidamping near the minimum.

An example of the `bare' flow curve, the final flow curve, and the region of instability is shown in Fig.1(a). Fig. 1(b) shows typical time series of the stress just inside, and well within, the unstable region. The limits of this region, $\sigma_c^\pm$, are
Hopf bifurcation points where there is onset of
finite frequency sinusoidal oscillations
with an amplitude varying as $|\gd-\gd_{c}|^{1/2}$.

Our choice of an exponential kernel is nongeneric: most integral kernels are not equivalent to any finite order differential equation \cite{Glendinning}. However, the above argument gives a generic mechanism of instability. If the flow curve is monotonic only because of a retarded term ($-\lambda<R'(\sigma)<0$), then temporal instability survives if the retardation time is too long. Its presence does not depend on details of the kernel, but  what it leads to might do so: in particular, chaos is impossible in a second-order system \cite{Glendinning} such as Eq.(\ref{three}). However, our finding of spontaneous oscillation but not chaos appears to be structurally stable: we were unable to find chaos with $M_2$ taken as the sum of two exponentials (which gives a third order dynamical system for which chaos is allowed).

In that case, what does need to be added to the model of
Eq.(\ref{one}) to give temporal chaos rather than just
spontaneous oscillation? So far, the simplest variant we
have found that definitely shows chaos is the following:

\begin{equation}
\dot\sigma(t) = \dot\gamma - R(\sigma_1) -\lambda\sigma_2
\label{six}
\end{equation}

\noindent{}where the stress in the nonlinear term,
$\sigma_1$, is now also retarded.
The steady state flow curve is the same as that for (\ref{one}).
For simplicity, we choose a single exponential kernel here too:
$\sigma_{1}(t)=\int_{0}^{t}\sigma(t')\tau_{1}^{-1}\exp[-(t-t')/\tau_1]\,{\rm d}t'$.
To maintain continuity
of interpretation with the simpler version of the model,
we choose
$\tau_1\stackrel{<}{\scriptstyle{\sim}}1\ll\tau_2$.
We study the situation where the monotonicity of the flow curve
(still given by Eq.(\ref{two})) is restored only via the more retarded of the two relaxation terms. While there is no longer a simple interpretation in terms of an effective potential or a damping function, the generic instability of the previous model remains. But now, within the unstable region, we find a period doubling cascade leading to chaos.
Figure 2(a) shows, for a specified set of model parameters,
the period and Lyapunov exponents
$\lambda_{1}\geq\lambda_{2}\geq\lambda_{3}$
as a function of $\dot\gamma$
($\lambda_{1}>0$ means that nearby trajectories
exponentially separate~\cite{Wolf});
Fig.2(b) shows a series of period-doubling orbits in the $(\sigma_2,\sigma)$ plane and Fig.2(c) shows the strange attractor in $(\sigma_1,\sigma_2,\sigma)$-space.
Its Lyapunov dimension
$D_{\rm lyap}=2+\lambda_{1}/|\lambda_{3}|$
varies with the parameters but is
slightly greater than 2,
typically $2.0<D_{\rm lyap}< 2.1$.

Physically it is not clear to us yet why retardation of the nonlinear term (as well as the linear one) seems necessary to get chaos out of Eq.(\ref{six}); presumably, however, this adds something which is missing even from the mesoscopic model of Ref.\cite{Head1} (where chaos remained absent despite the infinite order of the system). Attempts to associate the retarded stresses in this model with, say, higher moments of the distribution of local strains in the model of Ref.\cite{Head1} (where the first moment is the instantaneous stress) have so far proved unconvincing. A more detailed study is left for future work.

We conclude with a broader discussion. The key idea is that of a flow curve (for spatially homogeneous states) whose monotonicity is rescued only by a retarded contribution; if too retarded, this does not restore temporal stability because the system continues to amplify perturbations over short timescales. Although the equations involved will look rather different, very similar physics could arise in materials of shear-thinning type where shear banding is present \cite{Callaghan,Spenley,Berret} or narrowly avoided \cite{Milner}. It might be very interesting to look more closely in shear-thinning micellar systems where, by varying density and temperature, one can arrange a material whose flow curve is only just monotonic \cite{Berret}. Similar studies in colloids close to the transition from continuous to discontinuous shear thickening \cite{Laun} would also be valuable although this field is a lot less developed experimentally. 

Quite similar equations, but with different variables and interpretation, might describe a pre-existing shear banded flow, whose stability remains unclear in many cases \cite{Olmsted2}. The simplest scenario would ascribe a single coordinate to describe the bands (e.g. the position of the interface between them, assumed flat) and seek to develop equations for its time evolution. Chaotic behavior of such an interface, rather than of a spatially homogeneous stress, might be the explanation of rheochaos seen in various micellar systems\cite{Sood1}. In the case where one of the bands is a static gel, empirical models such as those proposed in Refs.\cite{Ajdari} have met with some success at explaining the observed (though not entirely steady \cite{Pine}) dependence of stress on strain rate when averaged across such a banded flow. Such models involve equations such as $\dot h = f(h) - 1/\sigma$ where $h$ is the width of a shear band, $f$ is a nonlinear term arising from the difference in concentration in the two bands and $\sigma$ is the stress \cite{Ajdari}. Under controlled strain rate conditions (say) $1/\sigma$ is linear in $h$ and the equation is not dissimilar to Eq.(\ref{one}) without retardation. If a slow process can be identified (possibly concentration equilibration), then a retarded version of this type of equation could share the generic instability of the models discussed above. 

Acknowledgements: MEC and DAH acknowledge the hospitality of the Institute for Theoretical Physics, UCSB, where part of this work was completed with the support of NSF Grant No. PHY99--07949.
AA acknowledges the hospitality and support of the University
of Edinburgh, where this work was initiated.



\begin{figure}
\centerline{\psfig{file=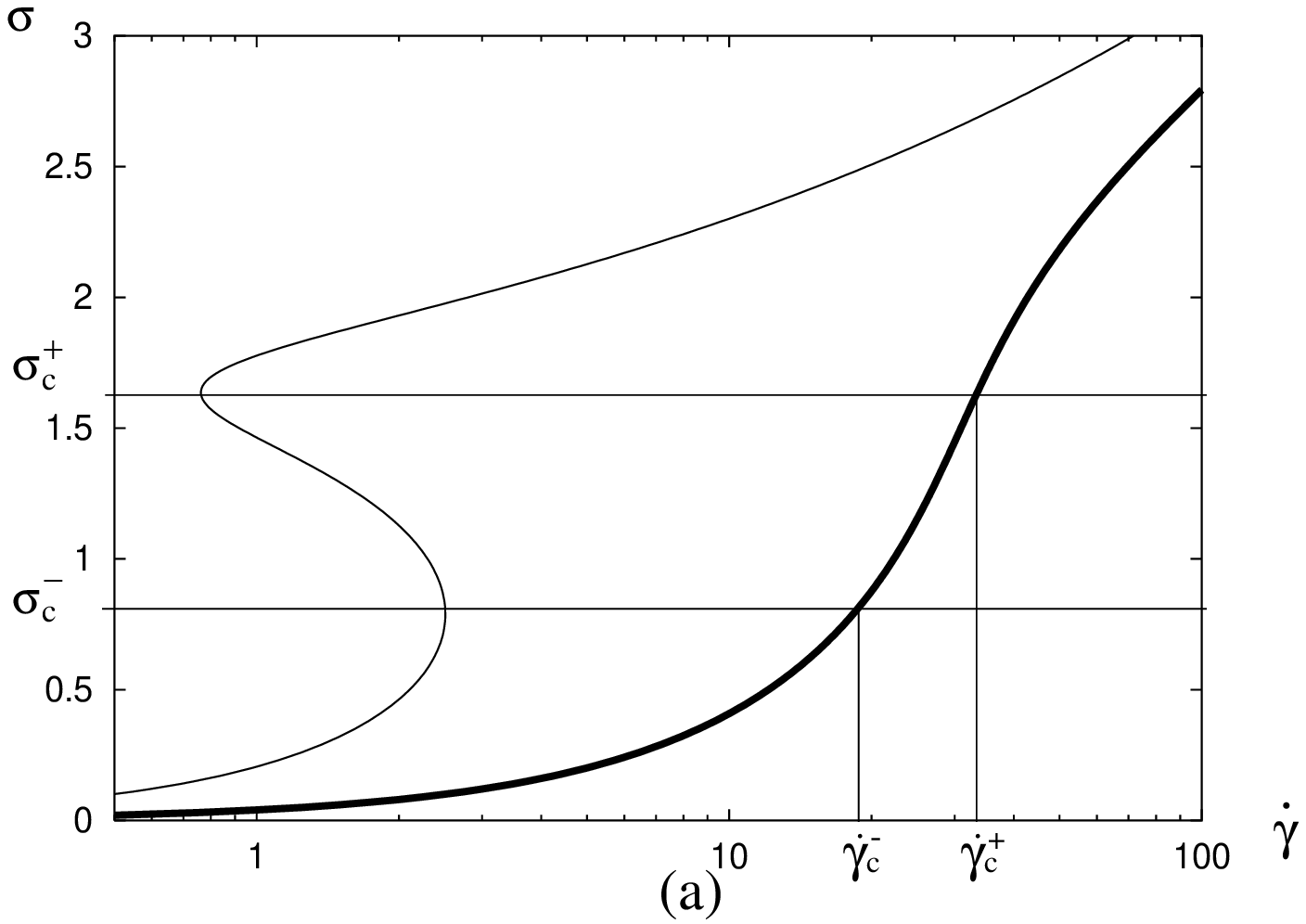,width=8cm}}
\centerline{\psfig{file=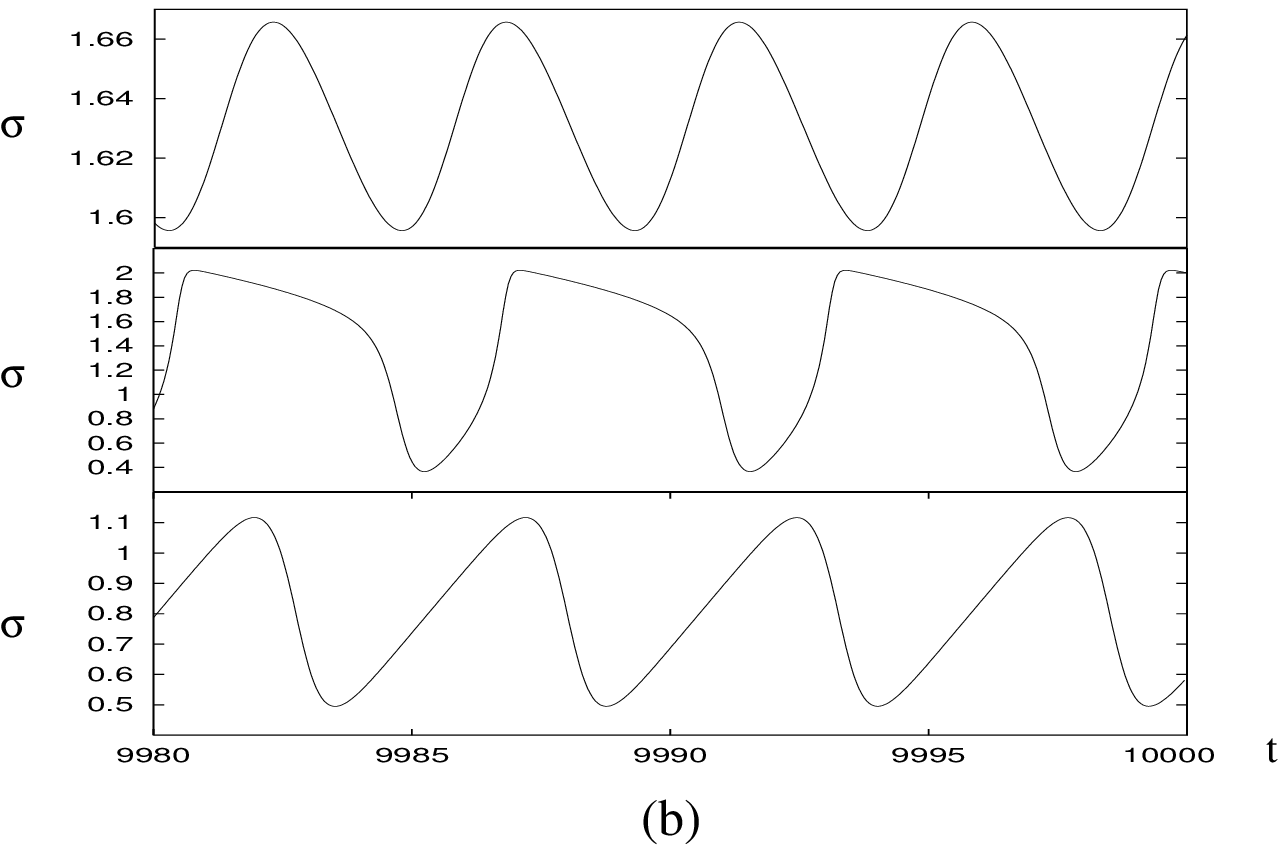,width=8.5cm}}
\caption{CAPTION: (a) The bare flow curve $\dot\gamma = R(\sigma)$ (light line) and the final flow curve (heavy line)
in the model of Eq.(\ref{one}).
Parameters are $\lambda=20$,
$\tau_{2}=10$ and 
$R(\sigma)=0.6\sigma^{5}-3.3\sigma^{3}+5\sigma$.
The region of instability
$\sigma_{\rm c}^{-}<\sigma<\sigma_{\rm c}^{+}$
is shown, where $\sigma_{\rm c}^{-}\approx0.799$
and $\sigma_{\rm c}^{+}\approx1.631$,
corresponding to
$\dot{\gamma}_{\rm c}^{-}\approx 18.487$ and
$\dot{\gamma}_{\rm c}^{+}\approx 33.382$.
Note that for our choice of parameters,
$\sigma^{\pm}_{\rm c}$ almost coincide with the
turning points of $R$.
(b) Stress time series (same parameter values)
at (from bottom to top)
$\dot{\gamma} = 18.49$, $\dot{\gamma} = 30$
and $\dot{\gamma} = 33.38$.
}
\end{figure}

\begin{figure}
\centerline{\psfig{file=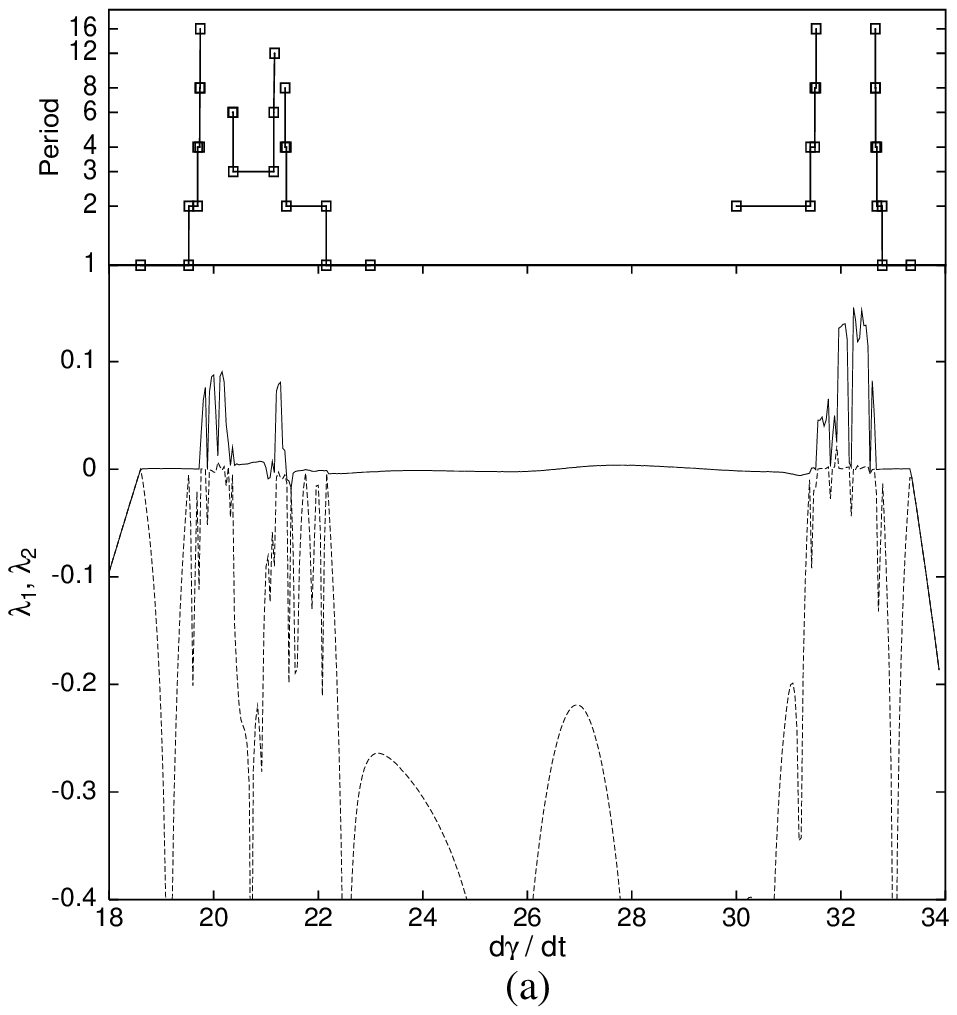,width=8.5cm,height=9cm}}
\centerline{\psfig{file=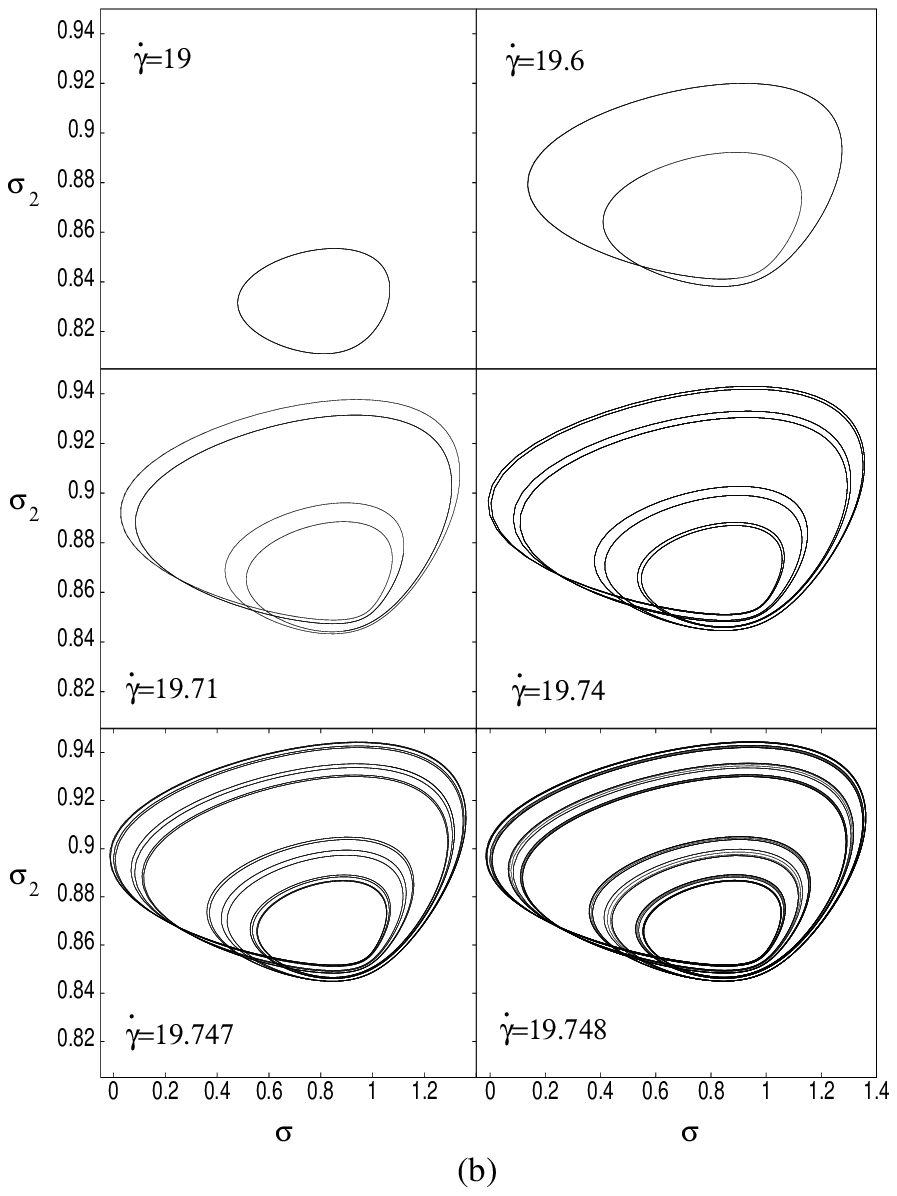,width=9cm,height=11cm}}
\centerline{\psfig{file=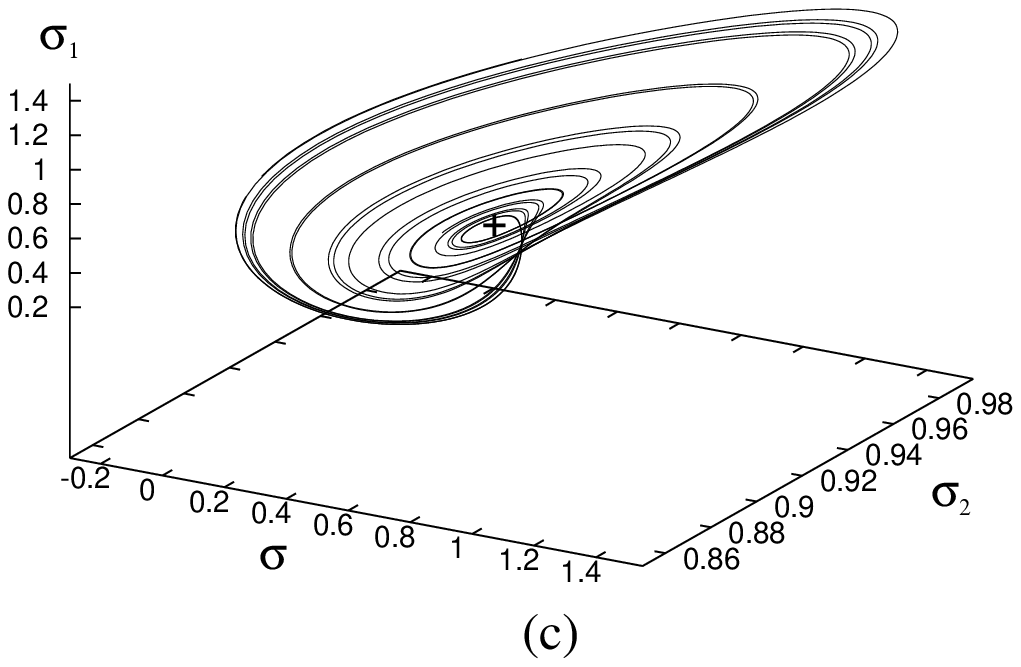,width=8cm}}
\caption{(a) Upper plot: the period of stable orbits as a function of strain rate $\dot\gamma$ around the unstable region
of the flow curve of Fig.1,
for the model of Eq.(\ref{six}) with
$\tau_{1}=0.5$ and the other parameters as in Fig.~1.
Lower plot:~Lyapounov exponents for trajectories, showing
$\lambda_{1}>0=\lambda_{2}$
in the chaotic regions.
(b) Orbits projected onto the ($\sigma_2,\sigma$) plane for various $\dot\gamma$ showing the period--doubling cascade
with periods 1, 2, 4, 8, 16 and 32.
(c) The strange attractor in ($\sigma_1,\sigma_2,\sigma$) space
for $\dot{\gamma}=20$ over a time period
$5\times10^{2}<t<10^{3}$ (arb. units).
}
\end{figure}

\end{multicols}


\begin{references}

\bibitem{Chaosfoot} Chaos is used here to mean bounded unsteady motion in a deterministic system that is neither periodic nor quasiperiodic, with trajectories that separate locally. We make no assumption about whether the chaos is low or high dimensional.

\bibitem{Chaikin} P. Chaikin in {\em Soft and Fragile Matter: Nonequilibrium Dynamics, Metastability and Flow},
ed. M.E. Cates and M.R. Evans
(IOP Publishing, Bristol, 2000).

\bibitem{Sood1} R. Bandyopadhyay and A.K. Sood,
Europhys. Lett. {\bf 56}, 447 (2002); R. Bandyopadhyay, G. Basappa and A.K. Sood,
Phys. Rev. Lett.~{\bf 84}, 2022 (2000).

\bibitem{Jacques} A.S. Wunenburger, A. Colin, J. Leng,
A. Arn«edo and D. Roux , Phys. Rev. Lett. {\bf 86},374 (2001); J.-B. Salmon, A. Colin and D. Roux, to be published.

\bibitem{Laun} H.M. Laun, J. Non--Newt. Fluid Mech.
{\bf 54}, 87 (1994); H. M. Laun, R. Bung and F. Schmidt, J. Rheol. {\bf 35}, 999 (1991); J. W. Bender and N. J. Wagner, J. Rheol. {\bf 40}, 899 (1996); W. J. Frith, P. d'Haene, R. Buscall and J. Mewis, J. Rheol {\bf 40}, 531 (1996); J. R. Melrose and R. C. Ball, Europhys. Lett {\bf 32}, 535 (1995); O. Hess and S. Hess, Physica A {\bf 207}, 517 (1994).

\bibitem{Callaghan} M. M. Britton and P. T. Callaghan, Eur. Phys. J. B {\bf 7}, 237 (1999); M. M. Britton, R. W. Mair, R. K. Lambert and P. T. Callaghan, J. Rheol. {\bf 43}, 897 (1999).

\bibitem{Groisman} A. Groisman and V. Steinberg,
Europhys. Lett. {\bf 43}, 165 (1998);
A. Groisman and V. Steinberg,
Nature {\bf 405}, 53 (2000).

\bibitem{Newstein} M.C. Newstein {\em et al.},
J. Chem. Phys. {\bf 111}, 4827 (1999).

\bibitem{Keunings} M. Grosso, R. Keunings, S. Crescitelli and
P.L. Maffettone, Phys. Rev. Lett. {\bf 86}, 3184 (2001);
G. Rien\"acker, M. Kr\"oger and S. Hess,
{\em ``Chaotic and regular shear--induced orientational
dynamics of nematic liquid crystals''},
in preparation.

\bibitem{Head1} D. A. Head, A. Ajdari and M. E. Cates,
Phys. Rev. E {\bf 64}, 061509 (2001);
D. A. Head, A. Ajdari and M. E. Cates,
Euro. Phys. Lett. {\bf 57}, 120 (2002).

\bibitem{Ball} T.C.B. McLeish and R.C. Ball, J. Polym. Sci. B
{\bf  24}, 1735 (1986).

\bibitem{Spenley} N.A. Spenley, M.E. Cates and T.C.B
Mcleish, Phys. Rev. Lett. {\bf 71}, 939 (1993).

\bibitem{Olmsted} P.D.Olmsted and C.Y.D. Lu,
Faraday Discuss {\bf 112}, 183 (1999).

\bibitem{Derec} C. Derec, A. Ajdari and F. Lequeux,
Eur. Phys. J. E  {\bf 4}, 355 (2001).

\bibitem{footnote} In this case, the flow curve is monotonic
for $R'(\sigma)+2\lambda\sigma>0$, but becomes
unstable when
$R'(\sigma)+\lambda\sigma+1/\tau_{2}<0$.

\bibitem{Glendinning} P. Glendinning,
{\em Stability, Instability, and Chaos
}(Cambridge University Press, Cambridge, 1994).

\bibitem{Wolf} A. Wolf, J.B. Swift, H.L. Swinney and
A. Vastano, Physica D {\bf 16}, 285 (1985).

\bibitem{Berret} J. F. Berret, G. Porte and J.P. Decruppe, Phys. Rev. E. {\bf 55}, 1688 (1997); J. Physique (France) II {\bf 4}, 1261 (1994).

\bibitem{Milner} A. E. Likhtman, S. T. Milner and T. C. B. McLeish, Phys. Rev. Lett. {\bf 85}, 4550 (2000).

\bibitem{Olmsted2} O. Radulescu and P.D. Olmsted,
J. Non-Newt. Fl. Mech. {\bf 91}, 141 (2000).

\bibitem{Ajdari} A. Ajdari, Phys. Rev. E {\bf 58}, 6294 (1998);
J.L. Goveas and D.J. Pine, Europhys. Lett.
{\bf 48}, 706 (1999).


\bibitem{Pine} Y.T. Hu, P. Boltenhagen and D.J. Pine,
J. Rheol. {\bf 42}, 1185 (1998).

\end{references}
\end{document}